\documentclass[conference]{IEEEtran}
\IEEEoverridecommandlockouts
\usepackage[nobreak,space,compress]{cite}
\usepackage[hidelinks]{hyperref}
\usepackage{amsmath,amssymb,amsfonts}
\usepackage{algorithmic}
\usepackage{graphicx}
\usepackage{textcomp}
\usepackage[most]{tcolorbox}
\usepackage{xcolor}
\usepackage{tikz}
\usepackage{url}
\def\BibTeX{{\rm B\kern-.05em{\sc i\kern-.025em b}\kern-.08em
    T\kern-.1667em\lower.7ex\hbox{E}\kern-.125emX}}
\begin{document}

\title{From First Use to Final Commit: Studying the Evolution of Multi-CI Service Adoption}

\author
{
  \IEEEauthorblockN{Nitika Chopra and Taher A. Ghaleb}
  \IEEEauthorblockA{Department of Computer Science\\
  Trent University\\
  Peterborough, ON, Canada\\
  \{nitikachopra,taherghaleb\}@trentu.ca
  }
}

\maketitle

\begin{abstract}
Continuous Integration (CI) services, such as GitHub Actions and Travis CI, are widely adopted in open-source development to automate testing and deployment. Though existing research often examines individual services in isolation, it remains unclear how projects adopt and transition between multiple services over time. To understand how CI adoption is evolving across services, we present a preliminary study analyzing the historical CI adoption of 18,924 Java projects hosted on GitHub between January 2008 and December 2024, adopting at least one of eight CI services, namely Travis CI, AppVeyor, CircleCI, Azure Pipelines, GitHub Actions, Bitbucket, GitLab CI, and Cirrus CI. Specifically, we investigate: (1) how frequently CI services are co-adopted or replaced, and (2) how maintenance activity varies across different services. Our analysis shows that the use of multiple CI services within the same project is a recurring pattern observed in nearly one in five projects, often reflecting migration across CI services. Our study is among the first to examine multi-CI adoption in practice, offering new insights for future research and highlighting the need for strategies and tools to support service selection, coordination, and migration in evolving CI environments.
\end{abstract}


\section{Introduction}
Continuous Integration (CI) services have become a foundational component of modern software development, enabling developers to automate code builds, testing, and deployment~\cite{Fowler_CI}. By streamlining integration tasks, CI services help teams reduce errors, accelerate development workflows, and improve overall software quality. Over the years, the CI ecosystem has grown significantly, with services such as Travis CI, GitHub Actions, CircleCI, GitLab~CI and AppVeyor offering a wide range of features.

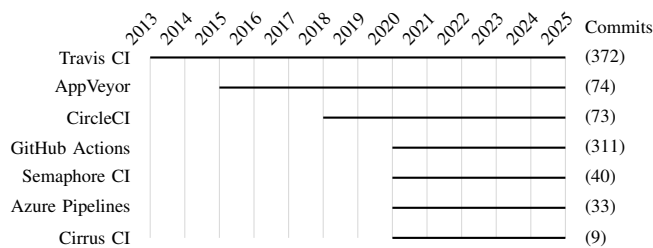
\begin{figure}[t]
\centering
\begin{tikzpicture}[xscale=0.46, yscale=1]

\foreach \x [evaluate=\x as \year using int(2013+\x)] in {0,...,12} {
  \node[anchor=north east, rotate=45] at (\x,0.1) {\scriptsize \year};
  \draw[gray!30] (\x,-0.7) -- (\x,-3.1);
}

\node[anchor=south west] at (12.3,-0.5) {\scriptsize Commits};

\draw[thick] (0,-0.7) -- (12,-0.7);  
\node[anchor=east] at (-0.3,-0.7) {\scriptsize Travis CI};
\node[anchor=west] at (12.3,-0.7) {\scriptsize (372)};

\draw[thick] (2,-1.1) -- (12,-1.1);  
\node[anchor=east] at (-0.3,-1.1) {\scriptsize AppVeyor};
\node[anchor=west] at (12.3,-1.1) {\scriptsize (74)};

\draw[thick] (5,-1.5) -- (12,-1.5);  
\node[anchor=east] at (-0.3,-1.5) {\scriptsize CircleCI};
\node[anchor=west] at (12.3,-1.5) {\scriptsize (73)};

\draw[thick] (7,-1.9) -- (12,-1.9);  
\node[anchor=east] at (-0.3,-1.9) {\scriptsize GitHub Actions};
\node[anchor=west] at (12.3,-1.9) {\scriptsize (311)};

\draw[thick] (7,-2.3) -- (12,-2.3);  
\node[anchor=east] at (-0.3,-2.3) {\scriptsize Semaphore CI};
\node[anchor=west] at (12.3,-2.3) {\scriptsize (40)};

\draw[thick] (7,-2.7) -- (12,-2.7);  
\node[anchor=east] at (-0.3,-2.7) {\scriptsize Azure Pipelines};
\node[anchor=west] at (12.3,-2.7) {\scriptsize (33)};

\draw[thick] (7,-3.1) -- (12,-3.1);  
\node[anchor=east] at (-0.3,-3.1) {\scriptsize Cirrus CI};
\node[anchor=west] at (12.3,-3.1) {\scriptsize (9)};

\end{tikzpicture}
\vspace{-15pt}
\caption{Timeline of CI service adoption in \texttt{checkstyle/checkstyle} (2013–2025).}
\vspace{-14pt}
\label{fig:ci_timeline}
\end{figure}

Despite the availability of numerous CI services, developers often face uncertainty in selecting and managing them throughout the project lifecycle~\cite{rostami2023usage,hilton2016usage}. The increasing diversity of CI services has introduced new complexities related to service adoption, configuration maintenance, and co-adoption patterns~\cite{rostami2023usage}. Prior research has primarily focused on the benefits of CI or on comparisons between services in isolation~\cite{gallaba2022lessons,elazhary2021uncovering}. However, there is limited understanding of the adoption, usage evolution, and maintenance of CI services in open-source projects.
Figure~\ref{fig:ci_timeline} gives an example of multi-CI adoption by the \texttt{checkstyle/checkstyle} project\footnote{\url{https://github.com/checkstyle/checkstyle}} on GitHub, clearly illustrating this complexity, having adopted seven different CI services between 2013 and 2025, including co-existing evolving configurations and that have been continuously co-maintained over 800 CI-related commits, with a median of 73 (mean is 130) commits per CI service. This example highlights the need for a deeper investigation of multi-CI adoption patterns in real-world development.

\vspace{1pt}
This paper makes the following contributions.
\vspace{-2pt}
\begin{itemize}
    \item We present the first empirical study analyzing CI adoption trends and patterns across eight services over 17 years, providing insights into their adoption, usage evolution, maintenance, and complexity.

    \item We make our dataset, scripts, and raw results/plots of historical multi-CI adoption publicly available~\cite{our_replication_package}, thus enabling researchers to explore further relevant factors, problems, and insights.

\end{itemize}

Prior research studied the historical CI adoption and migration practices of CI services. However, these studies solely rely on either using developer surveys~\cite{rostami2023usage} or examining a single CI service like CircleCI~\cite{gallaba2022lessons}.
In this paper, we present the first empirical study on the historical adoption practices of 18,924 Java projects hosted on GitHub between January 2008 and December 2024, and adopting at least one of eight CI services, namely Travis CI, AppVeyor, CircleCI, Azure Pipelines, GitHub Actions, Bitbucket, GitLab CI, and Cirrus CI.
Specifically, we examine how CI services are adopted, maintained, and co-utilized in GitHub-hosted Java projects.
We study how CI services differ in terms of adoption, usage evolution, configuration complexity, and maintenance activity. We also analyze the patterns of CI service co-adoption and switching.
Our empirical analysis reveals that while most projects rely on a single CI service, a large portion (nearly one in five) configure and run multiple CI services concurrently, with some projects using as many as six. This highlights a persistent, underreported practice of multi-CI adoption in real-world development.
Our findings aim to shed light on the practical realities of the multiple CI service adoptions in open-source development and to provide insights for both researchers and tool builders interested in improving CI integration strategies.

\vspace{2pt}
The rest of this paper is organized as follows. Section~\ref{sec:background_and_related_work} provides relevant background and reviews related work.
Section~\ref{sec:dataset} gives an overview of our data collection and processing.
Section~\ref{sec:evaluation} details the motivation, approach, and evaluation results of our research questions.
Section~\ref{sec:discussion} discusses our findings in practice.
Section~\ref{sec:threats_to_validity} highlights the validity threats.
Finally, Section~\ref{sec:conclusion} concludes the paper and suggests possible future work.

\section{Background and Related Work}
\label{sec:background_and_related_work}
    \subsection{CI Process and Services}  Continuous Integration (CI) is the practice of automatically building and testing code changes to detect issues early in the development cycle. In this study, eight widely used CI services are considered, each of which supporting automated build, test, and deployment workflows. \textit{GitHub Actions} is tightly integrated into GitHub and widely adopted across projects. \textit{Travis CI} was among the earliest popular tools, especially for open-source projects. \textit{CircleCI} provides scalable and flexible pipelines for various platforms. \textit{GitLab CI} offers built-in CI/CD capabilities within the \textit{GitLab} ecosystem. \textit{Bitbucket Pipelines} is native to Bitbucket Cloud and supports containerized builds. \textit{AppVeyor} targets cross-platform projects with strong Windows support. \textit{Azure Pipelines} supports a broad range of languages and platforms and integrates with multiple repositories. \textit{Cirrus CI} is a lightweight service with support for diverse operating systems and cloud environments.
    
    \subsection{Studies on CI adoption and configurations}
    A substantial body of research has examined Continuous Integration (CI) and Continuous Deployment (CD) practices in modern software development. Fowler's original best practices for CI adoption have shaped many empirical investigations into real-world implementation~\cite{Fowler_CI}. Hilton et al.~\cite{hilton2016usage} analyzed thousands of GitHub repositories and surveyed developers, finding that only around 40\% actively used CI, often due to limited experience and setup difficulties.

    Several studies highlighted the challenges developers face with CI. Widder et al.~\cite{widder2019conceptual} reported inconsistent tool behavior, complex configurations, and troubleshooting difficulties, while Elazhary et al. identified issues in automating UI tests and managing pull request workflows~\cite{elazhary2021uncovering}. Vassallo et al.~\cite{vassallo2020configuration} proposed tooling support for detecting configuration smells. Bouzenia and Pradel emphasized the high resource usage of GitHub Actions workflows~\cite{bouzenia2024resource}.

    Other studies explored pipeline performance and reliability. Jin and Servant evaluated trade-offs in strategies to improve CI effectiveness~\cite{jin2021helped}. Hilton et al. also studied over 22 million CircleCI builds, showing how instability and misconfiguration affect success rates~\cite{gallaba2022lessons}. Mazrae et al. conducted a qualitative study on CI service migration, noting a widespread shift from Travis CI to GitHub Actions due to usability and integration improvements~\cite{rostami2023usage}. Other studies investigated how configuration choices can impact build duration~\cite{ghaleb2019empirical,ghaleb2022studying} and build failures~\cite{ghaleb2019studying,ghaleb2022studying}.

    Although most prior studies focused on individual CI services or qualitative findings, our work provides one of the first large-scale empirical analyses of multi-CI adoption, co-adoption, and obsolescence. By analyzing nearly 19k Java projects over 17 years, we offer a quantitative view of how CI adoption evolves in practice.

\section{Data Collection and Processing}
\label{sec:dataset}
\vspace{-2pt}
To investigate the adoption patterns of Continuous Integration (CI) services, we conducted a large-scale empirical study of open-source Java projects hosted on GitHub. We chose Java because it is the most studied language in CI-related research, according to a recent systematic literature review~\cite{aidasso2025build} and mapping study~\cite{pando2023sms}. Our goal was to analyze how projects adopt, maintain, and potentially co-adopt multiple CI services over time, while also capturing aspects such as configuration complexity and developer engagement.
    
\subsection{Data Collection}
\label{sec:data_collection}
\vspace{-2pt}
We began by identifying active, non-fork Java repositories created between January 2008 and December 2024 with at least five GitHub stars to ensure relevance and quality. This resulted in a dataset of 122,062 projects. To detect CI service adoption, we analyzed the presence of configuration files and directories commonly associated with popular CI services (see Table~\ref{tab:ci_yml}), identifying 18,924 projects that adopted at least one CI service. Projects were tagged with the corresponding CI services based on the configurations found within their codebase, allowing us to capture both single and multi-service adoption per project. For each repository, we further collected metadata including commit history, contributing developers, and CI configuration files.

\begin{table}[h]
    \centering
    \vspace{-8pt}
    \caption{CI Services and Corresponding Configuration File Patterns}
    \vspace{-7pt}
    \begin{tabular}{l|l}
    \hline
    \textbf{CI Service} & \textbf{Configuration File Pattern} \\
    \hline
        GitHub Actions & \texttt{.github/workflows/*.yml} \\
        Travis CI & \texttt{.travis\.yml} \\
        CircleCI & \texttt{.circleci/config.yml} or \texttt{circle.yml} \\
        GitLab CI & \texttt{.gitlab-ci.yml} \\
        AppVeyor & \texttt{.appveyor.yml} or \texttt{appveyor.yml} \\
        Azure Pipelines~~~~~~~ & \texttt{azure-pipelines.yml} \\
        Bitbucket Pipelines & \texttt{bitbucket-pipelines.yml} \\
        Cirrus CI & \texttt{.cirrus.yml} \\
        \hline
    \end{tabular}
    \vspace{-8.3pt}
    \label{tab:ci_yml}
    \end{table}

\subsection{Characteristics Extraction}
\label{sec:characteristics}
We computed the following characteristics about projects and their adopted CI services.

\begin{itemize}
    \item \textbf{Adoption and Evolution:} History of CI adoption across services and among projects.

    \item \textbf{Configuration Complexity:} Lines of YAML configurations per CI service.

    \item \textbf{Maintenance Activity:} Ratio of CI configuration commits to total project commits..

    \item \textbf{Developer Engagement:} Ratio of unique developers contributing to CI configurations to the total number of developers in the project.

    \item \textbf{Co-adoption Patterns:} Frequency and combinations of CI services used concurrently.

    \item \textbf{Switching Behavior:} Time between the adoption of one CI service and the addition of another.

    \item \textbf{Abandoned/Obsolete Configurations:} CI services with missing or unmaintained configurations for a long time.
\end{itemize}

\subsection{Data Processing}
We used Python for all our data pre-processing, analysis, and visualization. To ensure consistency, we removed duplicates and normalized service identifiers where necessary.

\section{Evaluation Results}
\label{sec:evaluation}
This section presents our empirical analyses aimed at answering the research questions (RQs) defined in our study. For each RQ, we begin by outlining the motivation of the question and explaining why it is relevant to understanding CI service adoption. We then describe our methodological approach used to gain insights from our data. Finally, we report the key findings derived from our analyses, highlighting patterns, trends, and implications that emerged from the data.
Our replication package (data, scripts, and raw results) is publicly available on Zenodo~\cite{our_replication_package}.

\subsection{\textbf{RQ1: How do CI services differ from each other in terms of adoption, usage evolution, configuration complexity, and maintenance activity?}}

    \subsubsection{\textbf{Motivation}} Prior work has established the value of CI in software development, but there is limited understanding of how individual services differ in practice. With the growing diversity of CI services, developers must often choose among options that might offer similar functionality but differ in configuration effort, stability, and community adoption. This raises questions about how these services are adopted, maintained, and phased out over time in real-world projects.

    \vspace{2pt}
    \subsubsection{\textbf{Approach}} To explore this, we analyzed the 18,924 projects described in Section~\ref{sec:data_collection} using the characteristics extracted and explained in Section~\ref{sec:characteristics}.
    In particular, we performed our analyses on these characteristics as project-level metrics capturing CI service adoption lifespan, configuration complexity, and maintenance activity. Complexity was estimated using the number of lines in CI YAML files, and maintenance was computed as the ratio of CI-related commits to total project commits. This allowed us to compare the lifecycle and effort across CI services.
    We used the Kruskal–Wallis H-test\cite{kruskal1952use} to assess whether there are statistically significant differences between CI services. For post hoc comparisons, we conducted pairwise Mann–Whitney U tests~\cite{mann1947} with Benjamini–Hochberg correction~\cite{benjamini1995} to control the false discovery rate and identify which specific CI services differ significantly.
    
    \vspace{3pt}
    \subsubsection{\textbf{Results}} In this section, we report our findings about CI adoption and usage evolution over time across all CI services considered in this study.

\begin{figure}[ht]
            \centering
            \vspace{-10pt}
            \includegraphics[width=\linewidth]{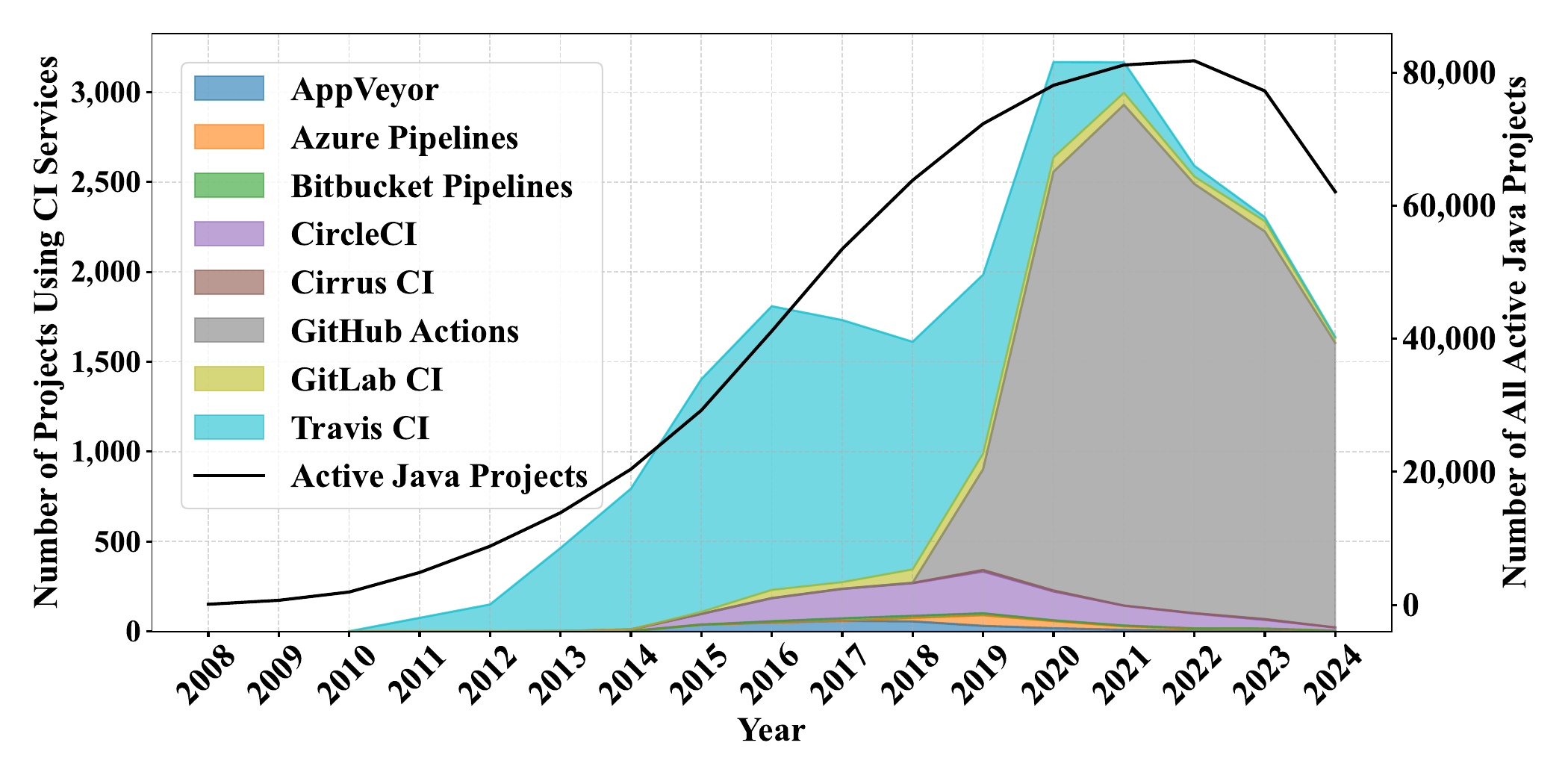}
            \vspace{-23pt}
            \caption{Stacked area chart showing the number of new Java GitHub projects adopting each CI service over time (2008–2024)}
            \vspace{-10pt}
            \label{fig:ci_adoption_trends}
        \end{figure}

        \vspace{5pt}
        \noindent\textbf{CI Service Adoption and Usage Evolution.}
        We found that 3,442 (approximately 18.15\%) of projects adopted two or more CI services, with 82\% of them having adopted both GitHub Actions and Travis CI. Some outlier projects used as many as six services simultaneously, likely due to legacy configurations, cross-platform requirements, or experimentation with newer tools. Figure~\ref{fig:ci_adoption_trends} shows the yearly adoption of CI services among Java projects on GitHub, alongside the number of all active Java repositories. CI adoption began rising sharply around 2013, with Travis CI leading the initial wave. However, Travis CI adoption declined significantly after 2016, likely due to the emergence of alternative services such as GitLab CI, GitHub Actions, CircleCI, and AppVeyor. Interestingly, Travis CI experienced a temporary rebound between 2018 and 2019, before dropping sharply again following its transition to a paid model in 2020. During this period, GitLab CI and GitHub Actions became two of the most widely adopted services.
        These results indicate that CI service adoption is dynamic, shaped by policy changes and developer preferences. The use of multiple CI setups and transitions highlights the need for better tools to handle CI diversity, facilitate service selection and migration, and reduce configuration redundancy.

\begin{figure}[htbp]
            \centering
            \vspace{-4pt}
            \includegraphics[width=\linewidth]{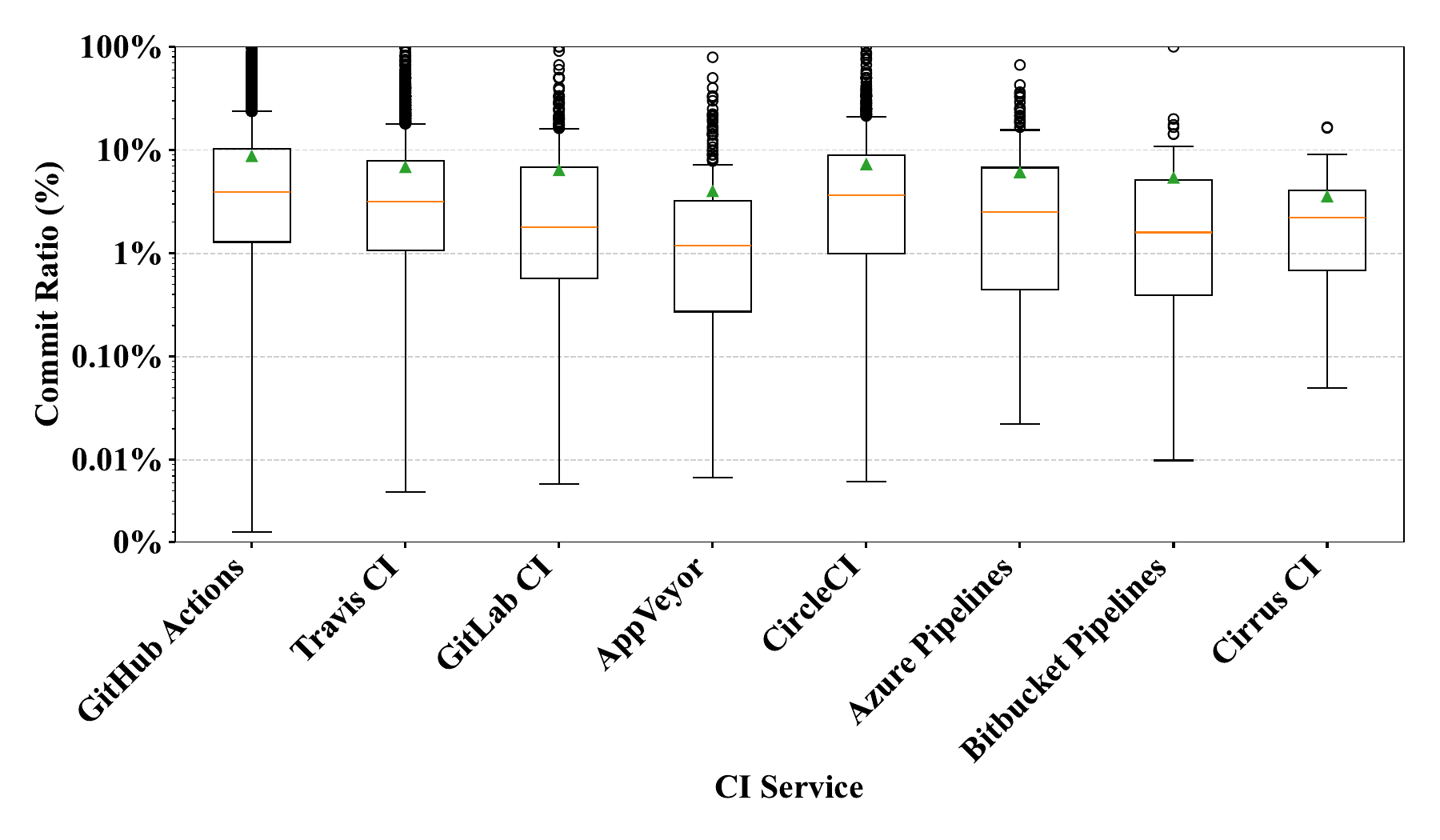}
            \vspace{-20pt}
            \caption{Box Plot chart shows the distribution of commits per project across CI services on a log scale.}
            \vspace{-6pt}
            \label{fig:ci_maintenance_activity}
        \end{figure}

        \vspace{5pt}
        \noindent\textbf{Maintenance Activity.}
        Figure~\ref{fig:ci_maintenance_activity} shows boxplots of CI maintenance activity over time for the eight CI services.
        We observe that almost all CI services show minimal CI maintenance activity. GitHub Actions was the highest with a median of only 3.9\% of total commits. This low proportion suggests that CI configurations are rarely updated relative to overall development activity. While that might indicate that CI workflows are stable, it might also raise concerns about neglect. Infrequent updates may lead to outdated, suboptimal, or even broken CI configurations, especially as project dependencies and workflows evolve. This highlights the need for better CI monitoring and tooling to prompt timely maintenance and reduce long-term risks~\cite{santos2025need}.

        \vspace{5pt}
        \noindent\textbf{Configuration Complexity.}
         Overall, all CI services have significantly different complexity from each other ($p-value <0.001$). Our results indicate that GitHub Actions tends to have a broader range in both file count and size, likely due to its ability to automate many repository processes beyond CI, resulting in more configurations. On the other hand, Travis CI shows relatively smaller YAML file sizes, implying a potentially simpler or more templated configuration style. The median YAML file line count for GitHub Actions was over $35$ lines, compared to approximately $20$ lines for Travis CI. GitLab CI exhibited the lowest variance, which might indicate consistent project-wide templates. These differences likely result from each CI service's configurability, documentation quality, or community norms.

        \begin{tcolorbox}[
              colback=gray!5,
              colframe=black,
              title=RQ1 Summary,
              boxsep=2pt,        
              left=4pt,          
              right=4pt,         
              top=2pt,           
              bottom=2pt         
            ]
            About 18\% of projects adopt multiple CI services, most commonly GitHub Actions and Travis CI, yet this likely comes with associated maintenance effort and configuration complexity.
        \end{tcolorbox}

    \subsection{\textbf{RQ2: What are the patterns of CI service co-adoption and switching among GitHub projects?}}
    
            \vspace{4pt}
            \subsubsection{\textbf{Motivation}} Although most open-source projects rely on a single CI service, a non-trivial subset appears to configure and maintain multiple services concurrently. This raises questions about the motivations behind co-adoption, the timing of transitions between services, and the presence of legacy or abandoned configurations. Existing literature often treats CI adoption as a static choice, overlooking the dynamic and sometimes unpredictable nature of CI co-adoption over a project's lifetime.

            \vspace{4pt}
            \subsubsection{\textbf{Approach}} To investigate these patterns, we analyzed the timing and order of multi-CI adoption across projects, identifying how quickly additional CI services were introduced and which combinations were most common.
                        In addition, we identified abandoned CI services by detecting repositories that had historically adopted a specific CI service, but no longer contain its configuration file in the current version.
                        To determine whether a CI service is obsolete, we checked if the repository has commits after the service's last configuration commit and if the inactivity period since that commit exceeds a given threshold (one year), ensuring the service is truly outdated despite ongoing project activity.

            \vspace{4pt}
            \subsubsection{\textbf{Results}} In this section, we report our findings about CI service abandonment, switching, and developer engagement in CI configuration activity.

        \vspace{3pt}
        \noindent\textbf{Abandoned and Obsolete CI Services.} Table~\ref{tab:ci_abandon_obsolete} summarizes the abandonment and obsolescence rates of various CI services, highlighting trends in their adoption and maintenance.
                We observe that roughly 18.8\% of CI configurations were eventually abandoned, while around 4.6\% became obsolete (rarely maintained) over time. GitHub Actions had the lowest abandonment (7.3\%) and obsolescence (1.5\%) rates, reflecting strong continued adoption. Travis CI had the highest number of outdated configurations, with moderate abandonment (29.3\%) and obsolescence (7.1\%) rates. CircleCI showed higher abandonment (41.3\%) and slightly higher obsolescence (7.7\%). Less widely adopted services, including GitLab CI, AppVeyor, Azure Pipelines, Bitbucket Pipelines, and Cirrus CI, exhibited high abandonment rates (28\% to 45\%) and varying obsolescence rates (7\% to 25\%). These results suggest that CI services with lower adoption or integration are more prone to being phased out, while newer or better integrated services maintain longer-term adoption.

        \begin{table}[ht]
        \centering
        \vspace{-4pt}
        \caption{Abandonment and Obsolescence of CI Services}
        \vspace{-5pt}
        \begin{tabular}{lrr}
            \hline
            \textbf{CI Service~~~~~~~~~~~~~~~} & \textbf{Abandoned \# (\%)} & \textbf{~~~~~~Obsolete \# (\%)} \\
            \hline
            Travis CI            & 2,573 (29.3\%) & 621 ~(7.1\%)  \\
            GitHub Actions       & 873 ~(7.3\%)   & 173 ~(1.5\%)  \\
            CircleCI             & 493 (41.3\%)   & 92 ~(7.7\%)   \\
            GitLab CI            & 151 (28.2\%)   & 70 (13.1\%)   \\
            AppVeyor             & 120 (44.6\%)   & 67 (24.9\%)   \\
            Azure Pipelines      & 60 (37.0\%)    & 21 (13.0\%)   \\
            Bitbucket Pipelines  & 24 (40.0\%)    & 10 (16.7\%)   \\
            Cirrus CI            & 9 (33.3\%)     & 2 ~~(7.4\%)   \\
            \hline
            \textbf{Overall}     & 4,303 (18.8\%) & 1,056 (4.6\%) \\
            \hline
        \end{tabular}
        \vspace{-5pt}
        \label{tab:ci_abandon_obsolete}
        \end{table}

        \vspace{3pt}
        \noindent\textbf{CI Switching.}
        Our analysis of 3,441 projects that switched CI services reveals that Travis CI is the predominant source of switches (71.5\%), while GitHub Actions is the primary target (74.4\%). Notably, 2,280 (26\%) of the projects that first adopted Travis CI switched to GitHub Actions. This is likely due to changes in service policies, such as Travis CI’s reduced free-tier support. Other common switches from other CI services to GitHub Actions also tend to happen more often, with ratios ranging from 23\% to 48\% of projects. In addition, we observe that the median time between switches is about 2.4 years (884 days), yet nearly 30\% of switches occur within one year.
        Overall, projects show a clear trend of migration to GitHub Actions, though the timing of such migrations varies widely.

        \vspace{3pt}
        \noindent\textbf{Developer Engagement in CI Configuration.} Surprisingly, we observe that 64\% of the studied projects have a single developer responsible for maintaining CI. 
        The median ratio of CI-maintaining developers to total contributors reinforces this pattern. GitHub Actions showed the highest typical ratio (38\%), followed by Travis CI and CircleCI (both at 33\%), suggesting a relatively greater involvement in these more commonly used services. In contrast, AppVeyor and Cirrus CI have lower medians (17\% and 24\%), indicating that CI maintenance is often handled by much fewer contributors. This concentration of responsibility can disrupt continuity, delay maintenance, and increase the risk that CI configurations become outdated or neglected.

        \begin{tcolorbox}[
          colback=gray!5,
          colframe=black,
          title=RQ2 Summary,
          boxsep=2pt,        
          left=4pt,          
          right=4pt,         
          top=2pt,           
          bottom=2pt         
        ]
        About 23\% of CI adoptions are later abandoned or become obsolete, often due to service switching, highlighting maintenance debt and the need for better migration and cleanup tools.
        \end{tcolorbox}

\section{Discussion}
~\label{sec:discussion}
Our findings raise several important points about how CI services are adopted, maintained, and phased out in practice. The key points are listed below.

\vspace{4pt}
\noindent\textbf{Multi-CI as a Strategic or Accidental Choice?}
Nearly one in five projects adopts multiple CI services, raising questions about whether developers intentionally combine services to meet specific needs or if these configurations arise from unmanaged drift. The presence of both Travis CI and GitHub Actions in the same project suggests that CI adoption might lack planning. This indicates a need to understand how CI decisions are made in open-source environments and why switching between CI systems occurs.

\vspace{3pt}
\noindent\textbf{CI Debt Is Silent and Overlooked.}
Our study reveals that about one-third of CI configurations are obsolete, including over 50\% of inactive Travis CI setups. These outdated configurations add to invisible CI debt by cluttering repositories and confusing contributors. They might also lead to redundant builds and waste CI resources, worsening infrastructure inefficiency. This supports earlier findings on technical debt of build systems~\cite{morgenthaler2012build}, highlighting underestimated long-term costs of unmanaged automation infrastructure.

\vspace{3pt}
\noindent\textbf{Configurability at a Cost.}
Our findings show that 18\% of projects adopt multiple CI services, with nearly half doing so within the first year, often switching from Travis CI to GitHub Actions. This challenges a common assumption, especially in the industry, that CI adoption is typically single-service and stable. In practice, developers balance configurability and platform-specific needs at the cost of increased maintenance complexity. The absence of support for CI migration or multi-CI coordination highlights the need for better infrastructure-aware development practices. This helps inform and shift community perceptions around real-world CI adoption.

\section{Threats to Validity}
\label{sec:threats_to_validity}
\vspace{-1pt}
    \subsection{Construct Validity}
    \vspace{-1pt}
        We identify CI service adoption by detecting configuration files in repositories, which may not always indicate active use, as some projects retain obsolete or unused files. Using the number of CI files as a proxy for configuration complexity might not capture the true semantic or logical complexity. Additionally, some projects may rely on external services or private repositories not visible in our dataset, potentially affecting the accuracy of metrics related to complexity or maintenance. Finally, since our analysis scripts were manually developed, human errors could impact the overall results.
        
    \subsection{Internal Validity}
    \vspace{-3pt}
    The timeline of CI adoption and switching was inferred from commit history, which may be affected by rebased or squashed commits, incomplete metadata, or repository restructuring, impacting accuracy. Developer counts based on GitHub usernames may not uniquely identify contributors due to aliasing or inconsistent naming. The analysis only captures visible CI adoption in commits, thus external or private configurations were excluded. Time gaps between commits may not reflect actual adoption or abandonment periods. Some CI transitions could be driven by organizational or policy changes not observable in the data. Further investigations are needed to assess these limitations and their potential impact on our conclusions.

    \subsection{External Validity}
    \vspace{-3pt}
        Our study focuses on Java projects on GitHub, which may limit the generalizability of our findings. CI adoption patterns could vary by programming language, application domain (e.g., data science, mobile apps, web apps), or services such as GitLab CI and Bitbucket. Therefore, caution is needed when applying these results outside this context.

\section{Conclusion}
~\label{sec:conclusion}
    In this paper, we conducted a preliminary empirical analysis of Continuous Integration (CI) service adoption patterns across 18,924 Java projects hosted on GitHub between January 2008 and December 2024, adopting at least one of eight CI services, namely Travis CI, AppVeyor, CircleCI, Azure Pipelines, GitHub Actions, Bitbucket, GitLab CI, and Cirrus CI. Our results reveal how the CI ecosystem has evolved over time, with early dominance by Travis CI gradually giving way to more modern and integrated solutions like GitHub Actions and GitLab CI. We observed that while most projects rely on a single CI service, a notable proportion adopt multiple CI services concurrently, sometimes retaining obsolete configurations long after switching. We also found that CI services have similarly minimal maintenance activity overall, and that the number of CI-contributing developers varies across CI services. These findings offer insight into how real-world projects adopt, evolve, and manage CI services in practice. Our results can inform future CI service development, developer onboarding strategies, and research on CI configuration maintenance. Future work may expand this investigation to other languages, explore CI-related failures or build outcomes, and leverage explainable models to recommend optimal CI strategies based on project characteristics.

\section*{Acknowledgment}
This research was supported by the Natural Sciences and Engineering Research Council (NSERC) Discovery Grant [RGPIN-2025-05897] and Trent University Knowledge Mobilization Research Grant.

\clearpage
\bibliographystyle{IEEEtran}
\bibliography{paper}

\begin{thebibliography}{10}
\providecommand{\url}[1]{#1}
\csname url@samestyle\endcsname
\providecommand{\newblock}{\relax}
\providecommand{\bibinfo}[2]{#2}
\providecommand{\BIBentrySTDinterwordspacing}{\spaceskip=0pt\relax}
\providecommand{\BIBentryALTinterwordstretchfactor}{4}
\providecommand{\BIBentryALTinterwordspacing}{\spaceskip=\fontdimen2\font plus
\BIBentryALTinterwordstretchfactor\fontdimen3\font minus \fontdimen4\font\relax}
\providecommand{\BIBforeignlanguage}[2]{{%
\expandafter\ifx\csname l@#1\endcsname\relax
\typeout{** WARNING: IEEEtran.bst: No hyphenation pattern has been}%
\typeout{** loaded for the language `#1'. Using the pattern for}%
\typeout{** the default language instead.}%
\else
\language=\csname l@#1\endcsname
\fi
#2}}
\providecommand{\BIBdecl}{\relax}
\BIBdecl

\bibitem{Fowler_CI}
M.~Fowler, ``{Continuous Integration},'' \url{https://martinfowler.com/articles/originalContinuousIntegration.html}, accessed: 2025-06-01.

\bibitem{rostami2023usage}
P.~Rostami~Mazrae, T.~Mens, M.~Golzadeh, and A.~Decan, ``On the usage, co-usage and migration of {CI/CD} tools: A qualitative analysis,'' \emph{Empirical Software Engineering}, vol.~28, no.~2, p.~52, 2023.

\bibitem{hilton2016usage}
M.~Hilton, T.~Tunnell, K.~Huang, D.~Marinov, and D.~Dig, ``Usage, costs, and benefits of continuous integration in open-source projects,'' in \emph{Proceedings of the 31st IEEE/ACM international conference on automated software engineering}, 2016, pp. 426--437.

\bibitem{gallaba2022lessons}
K.~Gallaba, M.~Lamothe, and S.~McIntosh, ``Lessons from eight years of operational data from a continuous integration service: An exploratory case study of {CircleCI},'' in \emph{Proceedings of the 44th international conference on software engineering}, 2022, pp. 1330--1342.

\bibitem{elazhary2021uncovering}
O.~Elazhary, C.~Werner, Z.~S. Li, D.~Lowlind, N.~A. Ernst, and M.-A. Storey, ``Uncovering the benefits and challenges of continuous integration practices,'' \emph{IEEE Transactions on Software Engineering}, vol.~48, no.~7, pp. 2570--2583, 2021.

\bibitem{our_replication_package}
N.~Chopra and T.~A. Ghaleb, ``From first use to final commit: Studying the evolution of multi-{CI} service adoption~({Replication Package}),'' \url{https://doi.org/10.5281/zenodo.16434046}, 2025.

\bibitem{widder2019conceptual}
D.~G. Widder, M.~Hilton, C.~K{\"a}stner, and B.~Vasilescu, ``A conceptual replication of continuous integration pain points in the context of {Travis CI},'' in \emph{Proceedings of the 2019 27th acm joint meeting on european software engineering conference and symposium on the foundations of software engineering}, 2019, pp. 647--658.

\bibitem{vassallo2020configuration}
C.~Vassallo, S.~Proksch, A.~Jancso, H.~C. Gall, and M.~Di~Penta, ``Configuration smells in continuous delivery pipelines: a linter and a six-month study on {GitLab},'' in \emph{Proceedings of the 28th ACM Joint Meeting on European Software Engineering Conference and Symposium on the Foundations of Software Engineering}, 2020, pp. 327--337.

\bibitem{bouzenia2024resource}
I.~Bouzenia and M.~Pradel, ``Resource usage and optimization opportunities in workflows of {GitHub Actions},'' in \emph{Proceedings of the 46th IEEE/ACM International Conference on Software Engineering}, 2024, pp. 1--12.

\bibitem{jin2021helped}
X.~Jin and F.~Servant, ``What helped, and what did not? an evaluation of the strategies to improve continuous integration,'' in \emph{2021 IEEE/ACM 43rd International Conference on Software Engineering (ICSE)}.\hskip 1em plus 0.5em minus 0.4em\relax IEEE, 2021, pp. 213--225.

\bibitem{ghaleb2019empirical}
T.~A. Ghaleb, D.~A. Da~Costa, and Y.~Zou, ``An empirical study of the long duration of continuous integration builds,'' \emph{Empirical Software Engineering}, vol.~24, pp. 2102--2139, 2019.

\bibitem{ghaleb2022studying}
T.~A. Ghaleb, S.~Hassan, and Y.~Zou, ``Studying the interplay between the durations and breakages of continuous integration builds,'' \emph{IEEE Transactions on Software Engineering}, vol.~49, no.~4, pp. 2476--2497, 2022.

\bibitem{ghaleb2019studying}
T.~A. Ghaleb, D.~A. Da~Costa, Y.~Zou, and A.~E. Hassan, ``Studying the impact of noises in build breakage data,'' \emph{IEEE Transactions on Software Engineering}, vol.~47, no.~9, pp. 1998--2011, 2019.

\bibitem{aidasso2025build}
H.~A{\"\i}dasso, M.~Sayagh, and F.~Bordeleau, ``Build optimization: A systematic literature review,'' \emph{arXiv preprint arXiv:2501.11940}, 2025.

\bibitem{pando2023sms}
B.~Pando and A.~D{\'a}vila, ``Software testing in the devops context: A systematic mapping study,'' \emph{Programming and Computer Software}, vol.~48, no.~8, pp. 658--684, 2022.

\bibitem{kruskal1952use}
W.~H. Kruskal and W.~A. Wallis, ``Use of ranks in one-criterion variance analysis,'' \emph{Journal of the American statistical Association}, vol.~47, no. 260, pp. 583--621, 1952.

\bibitem{mann1947}
H.~B. Mann and D.~R. Whitney, ``On a test of whether one of two random variables is stochastically larger than the other,'' \emph{The Annals of Mathematical Statistics}, vol.~18, no.~1, pp. 50--60, 1947.

\bibitem{benjamini1995}
Y.~Benjamini and Y.~Hochberg, ``Controlling the false discovery rate: a practical and powerful approach to multiple testing,'' \emph{Journal of the Royal Statistical Society: Series B (Methodological)}, vol.~57, no.~1, pp. 289--300, 1995.

\bibitem{santos2025need}
J.~Santos, D.~A. da~Costa, S.~McIntosh, and U.~Kulesza, ``On the need to monitor continuous integration practices,'' \emph{Empirical Software Engineering}, vol.~30, no.~5, p. 125, 2025.

\bibitem{morgenthaler2012build}
J.~D. Morgenthaler, M.~Gridnev, P.~Sauciuc, and S.~Bhansali, ``Searching for build debt: experiences managing technical debt at google,'' in \emph{Proceedings of the Third International Workshop on Managing Technical Debt}, 2012, pp. 1--6.

\end{thebibliography}

\end{document}